\documentclass[final,5p,times,twocolumn,authoryear]{elsarticle}
\usepackage{graphicx}
\usepackage{amsmath}
\usepackage[dvips]{epsfig}
\usepackage{subcaption}
\newcommand{\comm}[1]{}
\usepackage{hyperref}
\hypersetup{
  colorlinks=true,
  linkcolor=blue,
  citecolor=blue,
  urlcolor=red
  }

\begin{document}

\begin{frontmatter}

\title{Mass-Radius relationship of Strongly Magnetized Super-Chandrasekhar Anisotropic Deformed White Dwarf Stars in presence of $\gamma$-metric }

\author[first]{Rajasmita Sahoo}
\affiliation[first]{organization={National Institute of Technology, Tiruchirappalli, Tamil Nadu - 620015, India.}}

\ead{rsphysics58@gmail.com}

\author[second]{Tambe Pranjal Anant}
\affiliation[second]{organization={Inter-University Centre for Astronomy and Astrophysics, Pune, Maharashtra - 411007, India.}}

\ead{pranjaltambe32@gmail.com}

\author[first]{Somnath Mukhopadhyay}

\ead{somnath@nitt.edu}

\date{\today }

\begin{abstract}
The masses and radii of strongly magnetized anisotropic deformed white dwarf stars are investigated using the stellar structure equations in the parameterized $\gamma$-metric formalism. The Equation of State (EoS) of a completely degenerate relativistic electron gas in strong quantizing density-dependent magnetic field is developed. The fluid and field pressure anisotropy among the parallel and perpendicular components to the magnetic field is taken into consideration. This anisotropy in the EoS causes axisymmetric deformation of the star. We found stable solutions of deformed super-Chandrasekhar ultramassive white dwarfs. The masses of anisotropic magnetized white dwarfs at the same central density decrease monotonically with the increase in the strength of the central magnetic field, while the equatorial radii increase monotonically. This is in sharp contrast to the isotropic case where both the mass and radius increase monotonically. High magnetic field increases anisotropy and oblateness. We also see that the maximum mass and its corresponding equatorial radius both decrease as central magnetic field strength increases. We also notice that the maximum mass occurs at higher central density as the magnetic field increases. This shows that increasing magnetic field (hence increasing anisotropy) softens the EoS and makes the star more compact.

\end{abstract}


\end{frontmatter}


\noindent
\section{Introduction}
\label{Section 1}
The remnants of low- to medium-mass stars, like our Sun, that have exhausted nuclear fuel and are supported by electron degeneracy pressure are known as white dwarfs. In white dwarfs, the inward gravitational pressure is supported by the electron degeneracy pressure to maintain the star in hydrostatic equilibrium. The maximum mass of a nonmagnetized nonrotating white dwarf is the so called Chandrasekhar mass or Chandrasekhar limit \cite{Ch35} whose value is $\sim 1.44 M_{\odot}$. But recently observed Type Ia supernova, e.g., SN2006gz, SN2007if, SN2009dc, SN2003fg show exceptionally high luminosities that do not conform to the traditional Chandrasekhar limit \cite{Ho06,Sc10,Hi07,Ya09,Si11}. To account for these over-luminous Type Ia supernovae several models for the existence of stable white dwarfs beyond the Chandrasekhar limit, known as super-Chandrasekhar white dwarfs have been proposed. The presence of a strong internal magnetic field is one of the possibilities for supporting a massive super-Chandrasekhar white dwarf \cite{Mu17,Das12,Das13,Ni14,Roy19,Roy21}. In these studies, the electron gas in the EoS is taken to be free, relativistic and Landau quantized in a strong magnetic field. Several magnetized white dwarfs have been observed with surface field strengths ranging from $10^6$ to $10^9$ gauss \cite{Ke13,Ke15,Fe15}. Recently a highly magnetized rapidly rotating ultracompact white dwarf have been found with a mass of $\sim 1.3 M_{\odot}$ with a radius of about 2140 kms, a surface magnetic field of 600 megagauss and a rotation period of 6.94 minutes \cite{Ca21}.
The study of the physical properties of dense stellar matter in ultrahigh magnetic fields is an interesting and unsolved topic in theoretical astrophysics. In \cite{Das12} and \cite{Das13} the authors have used isotropic Landau quantized EoS (without magnetization) and anisotropic magnetic field pressures. They have used constant magnetic field and constant maximum Fermi energy to make the star into one, two and three Landau level system. They have found oblate spheroidal super-Chandrasekhar white dwarfs by solving Newtonian hydrostatic equilibrium equations. In \cite{Mu17}, the authors have used completely isotropic Landau quantized EoS of the free electron gas without magnetization. Also the magnetic field pressure is also taken to be isotropic (one-third of the energy density). Here the magnetic field is taken to be density-dependent inside the star, to make the matter thermodynamically stable. Hence there can be any number of Landau levels inside the star. The number of Landau levels depends on the radial distance from the centre as the number density and magnetic field both depend on the radial coordinate. By solving the Tolman-Oppenheimer-Volkoff (TOV) equations \cite{To39,Op39}, they have shown the existence of super-Chandrasekhar white dwarfs. Using this EoS, the r-mode instabilities and gravitational wave emission from magnetized white dwarfs have been found in \cite{Mu20}. In another work \cite{Roy19}, the authors have also incorporated the effects of Coulomb screening based on the Feynman-Metropolis-Teller treatment in presence of Landau quantization of electrons with density-dependent magnetic field profile. The EoS is considered isotropic in this case and stable super-Chandrasekhar white dwarf configurations have been found which depend on the elemental composition of the white dwarf. Using this isotropic EoS, the universal relations have been calculated in \cite{Roy21}. 
Recent studies on the properties of magnetized white dwarfs, neutron stars and quark stars using anisotropic magnetized EoS have been investigated in \cite{De21,De22}. The authors have taken both poloidal and toroidal magnetic field configurations with density-dependant magnetic field. They did not consider Landau quantization of the charged fermions, but considered the total anisotropy in the radial and transverse pressure as a smooth function of the radial coordinate. Then the masses and radii are found from the anisotropic TOV equations \cite{Bo74,He13}.
The deformation of magnetized compact stars due to the anisotropy in the matter and field EoSs breaks the spherical symmetry and introduces cylindrical symmetry in the system \cite{St12}. In \cite{Pa15,Par15,Te15}, the authors considered a metric in cylindrical coordinates to obtain Einstein’s field equations following the procedure described in \cite{Tr11}. However, due to their assumption that metric functions are only a function of the cylindrical polar radial coordinate  $r$, they were unable to obtain the total mass, but obtained the mass per unit length. In \cite{Zu17,Zub17}, the authors have developed a distorted Schwarzschild metric with a deformation parameter $\gamma$ that evaluates the degree of deviation from the spherical symmetry. This metric is called as $\gamma$-metric. The deformation parameter $\gamma$ is taken to be the ratio of the polar to equatorial radius of the star and is considered small. The mass of the star is calculated by solving the stellar structure equations in the radial and polar directions simultaneously. In \cite{Te19}, the first attempt has been made to solve the stellar structure equations using the '$\gamma$-metric' for magnetized white dwarfs with a constant magnetic field throughout the star. The deformation parameter $\gamma$ has been taken to be the ratio of the total parallel pressure to the total perpendicular pressure at the center of the star, and is taken to be constant throughout the entire star. This illustrates the relationship between the deformation and the anisotropy in the EoS. In their study, the authors have not found any star with mass greater than the Chandrasekhar limit. However, the use of a constant magnetic field may lead to thermodynamic instabilities, such as, the total parallel pressure may be negative when the field contribution is greater than the matter contribution. 
In this work, we have computed the masses and equatorial radii of magnetized white dwarfs with anisotropic EoS at absolute zero temperature in presence of $\gamma$-metric using the formulation of \cite{Te19} by taking into account density-dependent magnetic field profile within the star with variable Landau levels at every point inside the star and maintained surface magnetic field of $10^9$ gauss consistent with observations. The density-dependent magnetic field makes sure that the overall thermodynamic pressures (matter and field pressures) inside the star are positive definite and there isn't any thermodynamic instability. Using our density-dependant magnetized anisotropic EoS in the $\gamma$-metric formalism, we were able to find stable super-Chandrasekhar white dwarfs.
The anisotropic magnetized EoS for the electron gas is formulated in Section~\ref{Section 2}. The total anisotropic EoS of the white dwarf (matter and field contributions) field is formulated in Section~\ref{Section 3}. Sections~\ref{Section 4} and ~\ref{Section 5} are devoted to the formulation of the stellar structure equations in presence of $\gamma$-metric and the numerical procedure for solving the stellar structure equations with the anisotropic EoS. Sections~\ref{Section 6} and ~\ref{Section 7} discuss the results and summary and conclusions respectively.
\section{Anisotropic Equation of State for Magnetized Degenerate Electron Gas} 
\label{Section 2}
For the Anisotropic equation of state, we assume the magnetic field B to be pointing in the z-direction, so the electrons acquire discrete orbits in the plane perpendicular to the magnetic field B creating a quantization of the energy levels along the x- and y- directions (Landau quantization). The total energy of electrons is quantized into Landau levels, given by \cite{Mu17,St12},
\begin{equation}\label{eq:Energy}
    E_{\nu,p_{z}}=[{p_{z}^2 c^2}+m_{e}^2 c^4(1+2\nu B_{D})]^{1/2} .
\end{equation}
where $\nu = n + \frac{1}{2} + s_z$, $n = 0, 1, 2...$ is the Landau level number, and $s_z= {+\frac{1}{2},-\frac{1}{2}}$ is the z-component of spin of electron, $p_z$ is the z-component of linear momentum and $m_e$ is the electron rest mass. $B_D = \frac{B}{B_c}$ where $B_c = 4.414 \times 10^{13}$ Gauss is the critical magnetic field at which the cyclotron energy of electron equals its rest mass energy. The density of states (including spin degeneracy) in presence of magnetic field is given by
\begin{equation}\label{eq:dos}
 \sum_{\nu} \frac{2\pi}{h^2} m_e^2c^2 B_D g_{\nu} \int\limits \frac{dp_{z}}{h}.
\end{equation}
where, $g_\nu$ is the spin degeneracy of Landau levels, $g_\nu=1$ for $\nu=0$ and $g_\nu=2$ for $\nu\geq 1$.

As we are working in the zero temperature regime, the Fermi distribution function is,
\begin{equation}
f(E)=
\begin{cases}
    1, \; \mbox{for } E\leq E_F \\
    0, \; \mbox{for } E> E_F
\end{cases}
\end{equation}
where, $E_F$ is the Fermi-Energy, the maximum energy electron can occupy. This puts a constraint on $p_z$, with the maximum value ($p_{z,F}(\nu)$) given by,
\begin{equation}\label{eq:pzf}
p_{z,F}(\nu)= \frac{1}{c}\sqrt{{E_F}^2-m_{e}^2 c^4(1+2\nu B_{D})}.
\end{equation}
Thus we should have $\dfrac{{E_F}^2}{m_{e}^2 c^4}\geq (1+2\nu B_{D})$, which implies, 
$\nu\leq\nu_{max}$=$\dfrac{1}{2 B_D}\left(\dfrac{{E_F}^2}{m_{e}^2 c^4}-1\right)$, for a particular value of $B_{D}$.\\

The number density of electrons is thus \cite{Mu17,St12},
\begin{equation}\label{eq:nden}
\begin{split}
   n_e &=\dfrac{2\pi}{h^2} m_e^2c^2 B_D\sum_{\nu=0}^{\nu_{max}} g_{\nu} \int\limits_{-\infty}^{+\infty} f(E) \frac{dp_{z}}{h},\\
   &=\dfrac{4\pi}{h^2} m_e^2c^2 B_D\sum_{\nu=0}^{\nu_{max}} g_{\nu} \int\limits_{0}^{p_{z,F}(\nu)} \frac{dp_{z}}{h},\\
   &=\dfrac{4\pi}{h^3} m_e^2c^2 B_D\sum_{\nu=0}^{\nu_{max}} g_{\nu}\, {p_{z,F}(\nu)}.\\
\end{split}
\end{equation}
The energy density of electrons is given by \cite{Mu17,St12},
\begin{eqnarray}\label{eq:enden}
   \epsilon_e && =\dfrac{2\pi}{h^2} m_{e}^2c^2 B_D\sum_{\nu=0}^{\nu_{max}} g_{\nu} \int\limits_{-\infty}^{+\infty} f(E) E_{\nu,p_z} \frac{dp_{z}}{h},\nonumber \\
   =&&\dfrac{4\pi}{h^2} m_{e}^2c^2 B_D\sum_{\nu=0}^{\nu_{max}} g_{\nu} \times \nonumber
   \\&&\int\limits_{0}^{p_{z,F}(\nu)} \sqrt{{p_{z}^2 c^2}+m_{e}^2 c^4(1+2\nu B_{D})} \,\frac{dp_{z}}{h},\nonumber \\
   =&&\dfrac{2\pi}{h^3} m_{e}^2c^2 B_D\sum_{\nu=0}^{\nu_{max}} g_{\nu}\, \Bigg(E_F\; p_{z,F}(\nu) +  m_{e}^2 c^3 \times \nonumber
   \\&& (1+2\nu B_{D})  \log_e\left( \frac{E_F+p_{z,F}(\nu)c}{\sqrt{m_{e}^2 c^4(1+2\nu B_{D})}}\right) \Bigg) \nonumber.\\
\end{eqnarray}
Where we have used,
\begin{eqnarray}
\int \sqrt{x^2 + a^2}\, dx=\frac{1}{2}\Bigg(&&x\sqrt{x^2 + a^2} + \nonumber\\&& a^2 \log_e\left(\sqrt{\frac{x^2}{a^2}+1}+\frac{x}{a}\right) \Bigg)
.\nonumber
\end{eqnarray}
\\
The Anisotropic Parallel Pressure (z-direction) of electron gas in the direction of the magnetic field is given by  \cite{St12},
\begin{eqnarray}\label{eq:ppar}
   P_{\parallel e}=&& B_D\frac{2\pi m_e^2c^2}{h^3} \sum_{\nu=0}^{\nu_{max}} g_{\nu} \int\limits_{p_{z}=-\infty}^{+\infty} \frac{c^2 p_{z}^2}{E_{\nu,p_z}} f(E) dp_{z},\nonumber \\
  = && B_D\frac{4\pi m_e^2c^2}{h^3} \sum_{\nu=0}^{\nu_{max}} g_{\nu} \times \nonumber
  \\&& \int\limits_{0}^{p_{z,F}(\nu)} \frac{c^2 p_{z}^2}{\sqrt{{p_{z}^2 c^2}+m_{e}^2 c^4(1+2\nu B_{D})}} \, dp_{z},\nonumber\\
   = &&\dfrac{2\pi}{h^3} m_{e}^2c^2 B_D\sum_{\nu=0}^{\nu_{max}} g_{\nu}\, \Bigg(E_{F}\; p_{z,F}(\nu) -   \\&&
   m_{e}^2 c^3(1+2\nu B_{D})\, \log_e\left( \frac{E_{F}+p_{z,F}(\nu)c}{\sqrt{m_{e}^2 c^4(1+2\nu B_{D})}}\right) \Bigg)
  \nonumber.
\end{eqnarray}
Using, \\
$\int \frac{x^2}{\sqrt{x^2 + a^2}}\, dx=\frac{1}{2}\left( x\sqrt{x^2 + a^2} - a^2 \log_e\left(\sqrt{\frac{x^2}{a^2}+1}+\frac{x}{a}\right)\right)$.
\\
The Anisotropic Perpendicular  Pressure(x- and y- directions) of electron gas in the direction perpendicular to the magnetic field is given by \cite{St12},
\begin{eqnarray}\label{eq:pperp}
  P_{\perp e}=&& \frac{m_e^2c^4}{2}{\left(B_D\right)}^2\left(\frac{2\pi m_e^2c^2}{h^3}\right) \sum_{\nu=0}^{\nu_{max}} g_{\nu} \times \nonumber \\
  && \int\limits_{p_{z}=-\infty}^{+\infty} \frac{2\nu}{E_{\nu,p_{z}}} f(E) dp_{z},\nonumber\\
  =&& {m_e^2c^4}{\left(B_D\right)}^2\left(\frac{2\pi m_e^2c^2}{h^3}\right) \sum_{\nu=0}^{\nu_{max}} g_{\nu} \times \nonumber\\
  && \int\limits_{0}^{p_{z,F}(\nu)} \frac{2\nu}{\sqrt{{p_{z}^2 c^2}+m_{e}^2 c^4(1+2\nu B_{D})}}\,dp_{z},\nonumber\\
  =&& {m_e^2c^3}{\left(B_D\right)}^2\left(\frac{2\pi m_e^2c^2}{h^3}\right) \sum_{\nu=0}^{\nu_{max}} g_{\nu} \times \nonumber\\
  && 2\nu\, \log_e\left( \frac{E_{F}+p_{z,F}(\nu)c}{\sqrt{m_{e}^2 c^4(1+2\nu B_{D})}}\right).\nonumber 
\\
\end{eqnarray}
Using, $\int \frac{1}{\sqrt{x^2+a^2}}dx=\log_e\left(\sqrt{\frac{x^2}{a^2}+1}+\frac{x}{a}\right)$.
\section{Equation of State of Anisotropic Magnetized White Dwarfs}
\label{Section 3}
In anisotropic magnetized white dwarfs, the contribution of field energy density $\epsilon_{B}$=$\frac{B^2}{8\pi}$ and respective pressures $P_{\perp B}$=$\frac{B^2}{8\pi}$ and $P_{\parallel B}$=$-\frac{B^2}{8\pi}$ also come due to the presence of the anisotropy in the magnetic field itself \cite{Das12}. Now, we get the total energy density, total parallel pressure, and total perpendicular pressure of the white dwarf by adding both the fluid part and field part as follows, \\
\begin{equation}
\epsilon_{T}=\epsilon_{e}+n_e(m_p+m_n)c^2+\frac{B^2}{8 \pi},\label{eq:et}
\end{equation}
\begin{equation} \label{eq:pr}
    P_{\parallel T}=P_{\parallel e}-\frac{B^2}{8 \pi},
\end{equation}
\begin{equation} \label{eq:pn}
    P_{\perp T}=P_{\perp e}+\frac{B^2}{8 \pi}.
\end{equation}
The protons and the neutrons in magnetized white dwarfs contribute only a large amount of energy density to the system compared to degenerate electrons due to their higher mass and lower momentum. In Eq.~\eqref{eq:et}, the second term is the rest mass energy contribution of the protons and neutrons. $m_p$ and $m_n$ are the proton and neutron rest masses respectively. We have considered the number of neutrons to be the same as the number of protons.
\section{Stellar structure equations for Anisotropic Magnetized White Dwarfs}
\label{Section 4}
Due to the anisotropy in the parallel and perpendicular pressures, magnetized white dwarfs become axisymmetric and deformed to maintain hydrostatic equilibrium. To study the stellar structure equations of axisymmetric magnetized white dwarfs in the framework of general relativity, we construct a model based on \cite{Zu17,Zub17,Te19} to take account of the axisymmetric nature of the matter distribution and the spacetime around it (the $\gamma$-metric formalism). We assume that the magnitude of deformation is very less, so that the axisymmetric metric can be treated as a small perturbation to the spherical Schwarzschild metric. The line element for the deformed star is given as follows ($G=c=1$)
\begin{eqnarray}
ds^2=&&-e^{2\nu(r)}dt^2+\left(1-\frac{2m(r)}{r}\right)^{-\gamma}dr^2 \nonumber\\
&&+r^2 sin^2\theta d\phi^2+r^2d\theta^2. 
\end{eqnarray}
where $\nu(r)$ is a metric potential, $m(r)$ is the gravitational mass enclosed within radius $r$ and $\gamma$ is a parameter characterizing the degree of deformation. $\gamma$ depends on both the radial and polar coordinates such that $ \gamma = \frac{z}{r}$, where $r$ is the equatorial radius and $z$ is the polar radius. Our work is to show how the magnetized EoS affects the stellar structure equations and related anisotropy. So we have considered $\gamma$ as the ratio of total central parallel pressure to total central perpendicular pressure of the star and is taken to be a constant for a particular star with fixed central density and fixed central magnetic field strength.
\begin{equation}
    \gamma=\frac{P_{\parallel T0}}{P_{\perp T0}}.
\end{equation}
This assumption is now allowing us to relate the physics of the system with geometry. That means we can describe the star's shape and stability by the anisotropy of the EoS from the star’s center. So, the Einstein's field equations (EFE) are given by,
\begin{equation}
    G_{\mu\nu}=R_{\mu\nu}-\frac{1}{2} R g_{\mu\nu} = 8\pi T_{\mu\nu}
\end{equation}

Where $G_{\mu\nu}$ is Einstein's tensor, $R_{\mu\nu}$ is the Ricci curvature tensor, R is the Ricci scalar, $g_{\mu\nu}$ is the metric tensor and $T_{\mu\nu}$ is the energy-momentum tensor.

 The stellar structure equations resulting from the  Einstein's field equations using the metric described above and the energy-momentum tensor derived from the anisotropic magnetized EoS, are presented below \cite{Te19}: 
\begin{eqnarray}\frac{dm}{dr}= 4\pi r^2 \gamma \epsilon_T \nonumber\\   \label{eq:tov}
\frac{dP_{\parallel T}}{dz}=-\frac{\left(\epsilon_T+P_{\parallel T}\right)\left[\frac{r}{2}+4\pi r^3 P_{\parallel T}-\frac{r}{2}\left(1-\frac{2m}{r}\right)^\gamma\right]}{\gamma r^2 \left(1-\frac{2m}{r}\right)^\gamma} \nonumber \\
\frac{dP_{\perp T}}{dr}=  -\frac{\left(\epsilon_T+P_{\perp T}\right)\left[\frac{r}{2}+4  \pi r^3 P_{\perp T}-\frac{r}{2} \left(1-\frac{2m}{r}\right)^\gamma\right]}{ r^2 \left(1-\frac{2m}{r}\right)^\gamma}\nonumber \\
\end{eqnarray}

The solutions of the above stellar structure equations can be numerically computed just like the spherically symmetric TOV equations. We start from the centre of the star by taking a value of the central number density. From the EoS, we get the central values: $\epsilon_{T0}=\epsilon_T(r=0)$,  $P_{\parallel T0} =  P_{\parallel T} (r=0)$, $P_{\perp T0} = P_{\perp T} (r=0)$ and $\gamma=P_{\parallel T0}/P_{\perp T0}$, which is kept constant throughout the star. We stop the computation when the lesser of the two pressures, in this case $P_{\parallel T}$, goes to zero as mentioned in \cite{Te19}. This gives the value of the polar radius $Z$ of the star, $P_{\parallel T}(Z)=0$. From $Z$ we can find the equatorial radius $R$ as $R=Z/\gamma$. The total mass of the star is given as $M=m(R)$. It's worth mentioning that by setting B = 0, the model will automatically return $P_{\perp}$ = $P_{\parallel}$ and $\gamma$ =1. This implies that we can get the spherically symmetric TOV equations from Eq.~\eqref{eq:tov}, which is the standard nonmagnetic solution for compact objects.

\section{Theoretical Calculations}
\label{Section 5}

As discussed in \cite{Te19}, from each integration of the parallel and perpendicular pressure differential equations from Eq.~\eqref{eq:tov}, we get two values of energy density after interpolation in every step. This means that mass distribution also becomes anisotropic. If we want to incorporate the anisotropy in the mass (energy) density in the stellar structure equations we can write it in the following form \cite{Te19}

\begin{eqnarray}
\frac{dm}{dr}= 4\pi r^2 \gamma \left( \frac{\epsilon_{\parallel T}+ \epsilon_{\perp T}}{2}\right) \label{eq:mtov} \nonumber\\   
\frac{dP_{\parallel T}}{dz}=-\frac{\left(\epsilon_{\parallel T}+P_{\parallel T}\right)\left[\frac{r}{2}+4\pi r^3 P_{\parallel T}-\frac{r}{2}\left(1-\frac{2m}{r}\right)^\gamma\right]}{\gamma r^2 \left(1-\frac{2m}{r}\right)^\gamma} \nonumber \\
\frac{dP_{\perp T}}{dr}=  -\frac{\left(\epsilon_{\perp T}+P_{\perp T}\right)\left[\frac{r}{2}+4  \pi r^3 P_{\perp T}-\frac{r}{2} \left(1-\frac{2m}{r}\right)^\gamma\right]}{ r^2 \left(1-\frac{2m}{r}\right)^\gamma}\nonumber \\
\end{eqnarray}

By taking the average energy density, we update the right hand of Eq.~\eqref{eq:mtov} without losing any information regarding the mass density anisotropy. We have solved Eq.~\eqref{eq:mtov} numerically with the previously mentioned boundary conditions to get the masses and radii of anisotropic magnetized white dwarfs with density-dependent magnetic field for fixed central electron number densities and central magnetic field strengths. The mass density for white dwarfs can be expressed in units of $2 \times 10^9$ gms/cc by multiplying number density of electrons $n_e$ expressed in units of $fm^{-3}$ by the factor $1.6717305 \times 10^6$.

The magnetic field profile as a function of the number density inside the white dwarf is taken to be of the following form \cite{Ba97}

\begin{equation} \label{eq:B}
    B_D = B_S + B_0[1-exp(-\alpha(n_e/n_0)^{\beta})]
\end{equation}
where $B_D$ (in units of $B_c$) is the magnetic field at electronic number density $n_e$, $B_S$ (in units of $B_c$) is the surface magnetic field, $n_0$ is the central electron number density and
$\alpha$, $\beta$ and $B_0$ are constants. We choose $\alpha=0.8$, $\beta=0.9$ and the surface magnetic field strength of $10^9$ Gauss, consistent with observations. Once we fix the central magnetic field strength, we can get the value of $B_0$. We have kept the upper limit of central magnetic field strength at $8B_c$ which is $3.5312 \times 10^{14}$ gauss, within the lower of the maximum limit suggested by N. Chamel et al. \cite{Ch13}.
\noindent
\section{Results and Discussion }
\label{Section 6}
\begin{figure}[htbp]
\vspace{0.0cm}
\eject\centerline{\epsfig{file=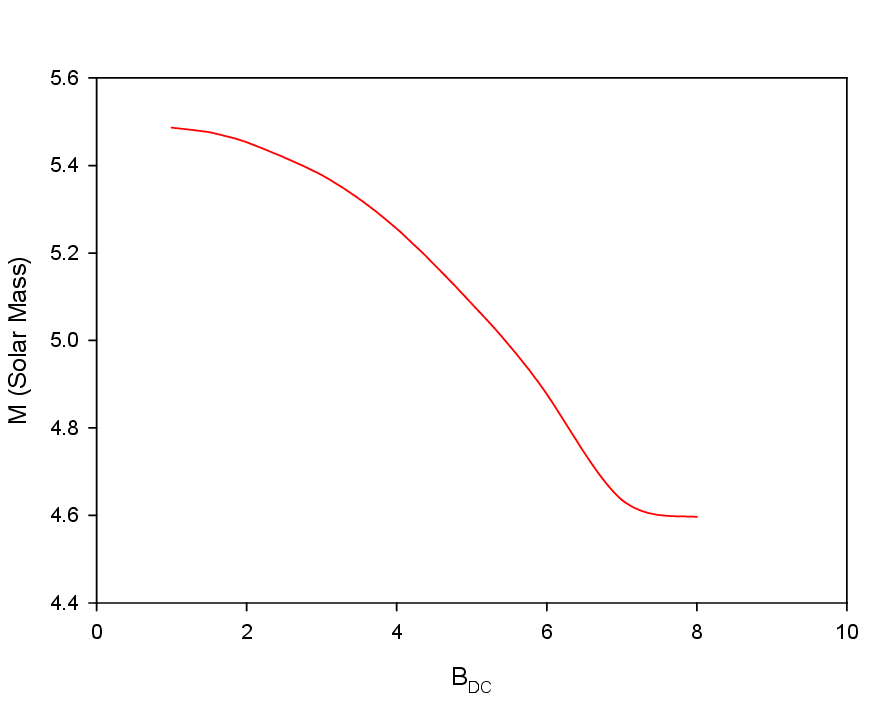,height=9cm,width=9cm}}\caption{Plot of Mass as a function of central magnetic field strength of anisotropic magnetized white dwarfs for a fixed central electron density $\sim 2.7 \times 10^{-6}$ $fm^{-3}$ .}
\label{fig1}
\vspace{0.0cm}
\label{fig:MB}
\end{figure}

\begin{figure}[htbp]
\vspace{0.0cm}
\eject\centerline{\epsfig{file=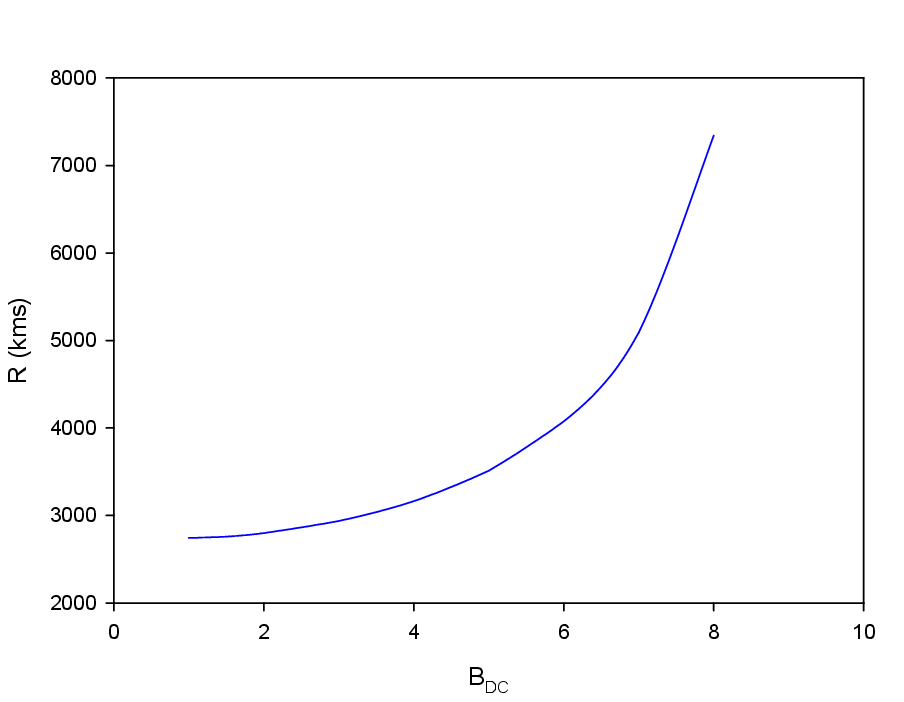,height=9cm,width=9cm}}
\caption{Plot of Equatorial Radius as a function of central magnetic field strength for anisotropic magnetized white dwarfs for a fixed central electron density $\sim 2.7 \times 10^{-6}$ $fm^{-3}$.}
\vspace{0.0cm}
\label{fig:RB}
\end{figure}

Here results for the masses and equatorial radii of anisotropic magnetized deformed white dwarfs are computed in presence of $\gamma$-metric, after performing the numerical calculations of the stellar structure equations Eq.~\eqref{eq:mtov} using the anisotropic EoS (Eqs.~\eqref{eq:et},~\eqref{eq:pr} ,~\eqref{eq:pn}) with the density-dependent magnetic field profile Eq.~\eqref{eq:B}. Figs.~\hyperref[fig:MB]{\ref*{fig:MB}} and ~\hyperref[fig:RB]{\ref*{fig:RB}} show the plots of the masses and equatorial radii vs. the dimensionless central magnetic field ($B_{DC}$) of anisotropic magnetized white dwarfs for a fixed central electron number density of  $\sim 2.7 \times 10^{-6}$ $fm^{-3}$ to show the effect of anisotropy due to high magnetic field. We found stable configurations of super-Chandrasekhar masses with density-dependent magnetic field profiles. The same values are listed in Table~\hyperref[table1]{\ref*{table1}}.

\begin{table}[ht]
\centering
\caption{Variations of masses and equatorial radii of anisotropic magnetized white dwarfs with dimensionless central magnetic field $B_{DC}$ at a fixed central electron number density of $\sim 2.7 \times 10^{-6}$ $fm^{-3}$ . The surface magnetic field $B_S$ is taken to be 10$^{9}$ gauss. The central electron number density can be expressed in units of 2$\times10^9$ gms/cc for mass density by multiplying with 1.6717305$\times10^6$.}
\begin{tabular}{||c|c|c||}
\hline 
\hline
$~~~~$$B_{DC}$$~~~~$&$~~~~$Mass $(M_\odot)$$~~~~$&$~~~~$Equatorial Radius (Kms)$~~~~$ \\ \hline
\hline
1&5.4865&2742.4984\\
2&5.4533&2798.5847 \\
3&5.3782&2937.1843 \\
4&5.2555&3162.3636 \\
5&5.0840&3511.1185 \\
6&4.8772&4075.0246 \\ 
7&4.6353&5090.9228 \\  
8&4.5965&7343.2344 \\  \hline
\hline
\end{tabular}
\label{table1} 
\end{table}
\noindent 
 

\begin{figure}[htbp]
  \vspace{0.0cm}
\eject\centerline{\epsfig{file=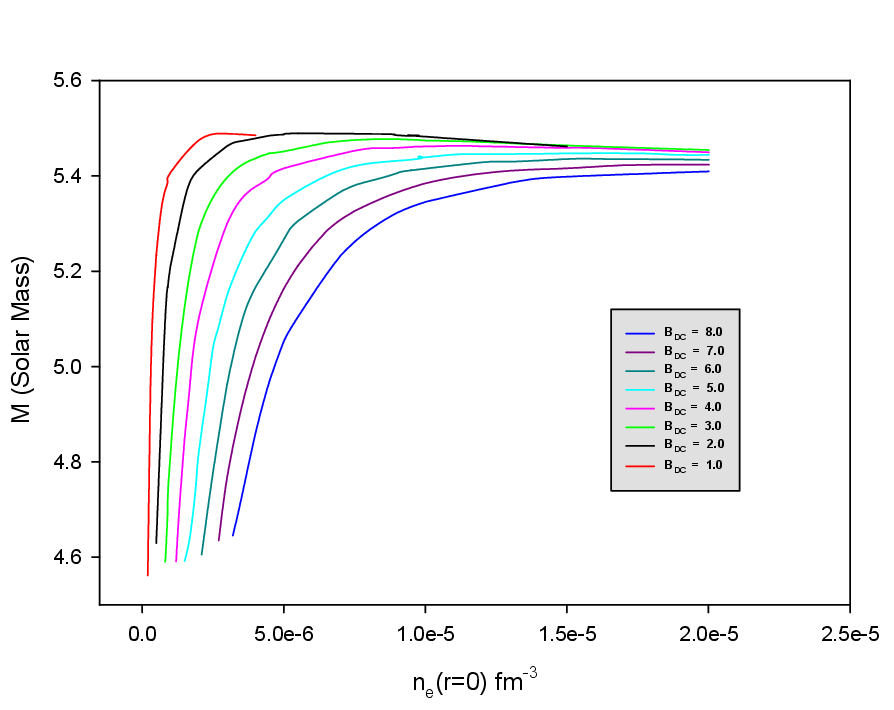,height=9cm,width=9cm}}
 \caption{Mass as a function of central electron number density for $B_{DC}$= 1 to 8.}
 \label{fig:MRhoce}
\end{figure}
\noindent

From Figs.~\hyperref[fig:MB]{\ref*{fig:MB}}, \hyperref[fig:RB]{\ref*{fig:RB}} and Table ~\hyperref[table1]{\ref*{table1}} we see that the mass goes on decreasing monotonically with the increase of dimensionless central magnetic field strength $(B_{DC})$, whereas the equatorial radius increases monotonically with increasing $B_{DC}$.  This result is in contrast to that of the isotropic case \cite{Mu17} where both the mass and the radius increase monotonically with $B_{DC}$. This variation can be attributed to the fact that high magnetic fields induce high anisotropy which thereby softens the EoS. When the magnetic field strength is very high, the total parallel pressure $P_{\parallel T}$ decreases drastically due to the increase of the field pressure (second term in Eq.~\eqref{eq:pr}). On the other hand the contribution of the field to the total perpendicular pressure $P_{\perp T}$ increases (second term in Eq.~\eqref{eq:pn}).  But in high magnetic fields the perpendicular pressure due to the electron gas $P_{\perp e}$ decreases due to lower number of Landau level occupation (Eq.~\eqref{eq:pperp}). Hence the total perpendicular pressure $P_{\perp T}$ does not increase much drastically. A more convenient way is to think of an effective isotropic average pressure $P_{avg}=(P_{\parallel T}+2P_{\perp T})/3$. At a fixed number density, this average pressure reduces as magnetic field strength increases, implying softening of the EoS and decrease of mass. 

\begin{figure}[htbp]
  \vspace{0.0cm}
\eject\centerline{\epsfig{file=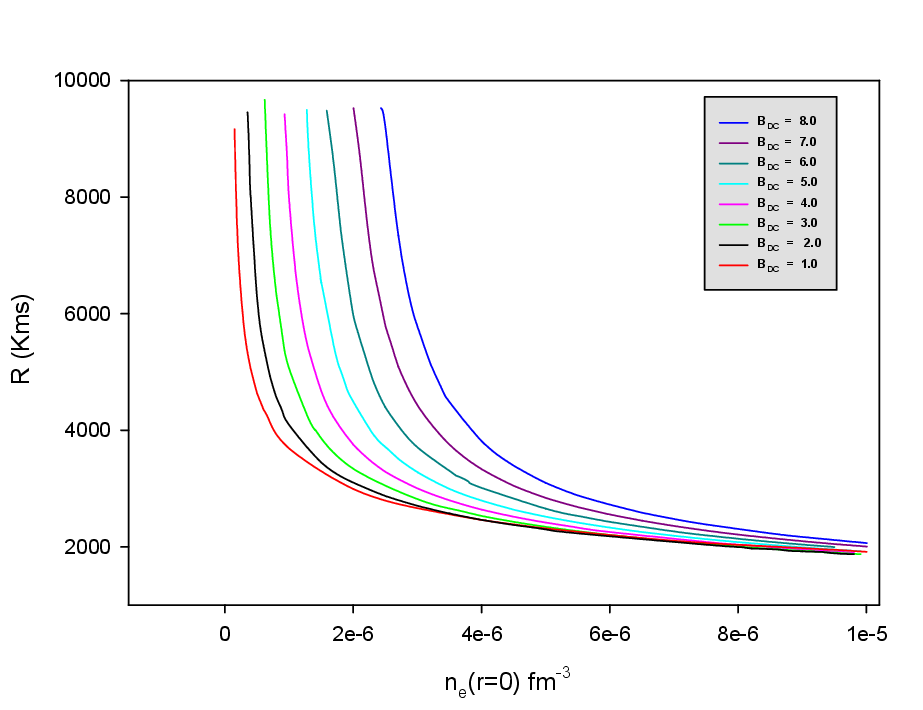,height=9cm,width=9cm}}
  \caption{ Equatorial radius as a function of central electron number density for $B_{DC}$= 1 to 8.}
 \label{fig:RRhoce}
 \end{figure}

\begin{figure}[htbp]
  \vspace{0.0cm}
\eject\centerline{\epsfig{file=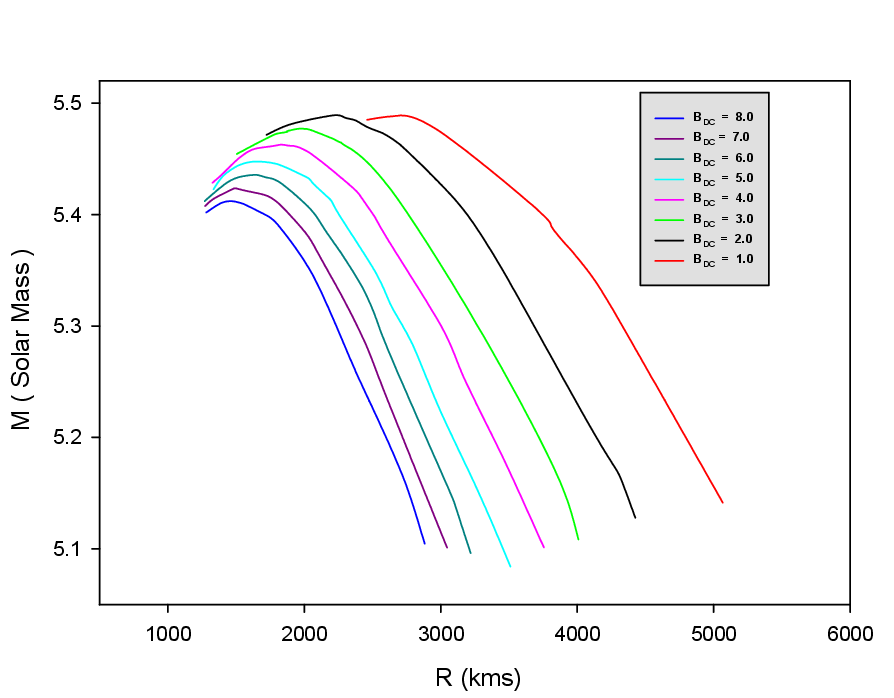,height=9cm,width=9cm}}
    \label{fig:figure3a}
    \label{fig:figure3b}
  \caption{ Mass vs. Equatorial radius for $B_{DC}$= 1 to 8.}
 \label{fig:MR}
\end{figure}

Figs.~\hyperref[fig:MRhoce]{\ref*{fig:MRhoce}},~\hyperref[fig:RRhoce]{\ref*{fig:RRhoce}} and ~\hyperref[fig:MR]{\ref*{fig:MR}} show the plots of Mass vs. Central electron number density, Equatorial radius vs. Central electron number density and Mass vs. Equatorial radius of anisotropic magnetized white dwarfs for $B_{DC}=1,2,3,4,5,6,7$ and $8$  respectively. 

From Fig.~\hyperref[fig:MRhoce]{\ref*{fig:MRhoce}}, the maximum mass value is found to decrease and the maximum mass occurs at a higher central electron number density when the magnetic field strength is increased. For $B_{DC}=1$, the maximum mass is found to be $5.4884 M_{\odot}$ at $n_e=3 \times 10^{-6} fm^{-3}$. For $B_{DC}=8$, the maximum mass is found to be $5.4121 M_{\odot}$ at $n_e=2.4 \times 10^{-5} fm^{-3}$. 
\noindent 
From Fig.~\hyperref[fig:RRhoce]{\ref*{fig:RRhoce}}, we find that for $B_{DC}=1$ the equatorial radius corresponding to the maximum mass at $n_e=3 \times 10^{-6} fm^{-3}$ is 2662.2390 kms and for $B_{DC}=8$ the equatorial radius corresponding to the maximum mass at $n_e=2.4 \times 10^{-5} fm^{-3}$ is 1457.9595 kms. This shows that the equatorial (and polar) radii decrease corresponding to the maximum masses as $B_{DC}$ increases, since they are occuring at higher central electron number densities. This clearly shows that increasing magnetic field strength leads to increasing compactness of white dwarfs due to the anisotropic softening of the EoS. The above fact can also be verified from the mass-equatorial radius plot (Fig.~\hyperref[fig:MR]{\ref*{fig:MR}}).
\noindent

\section{Summary and Conclusions }
\label{Section 7}
In this work, we have formulated an anisotropic EoS of a completely degenerate relativistic electron gas at absolute zero temperature in a strong quantizing magnetic field. The thermodynamic pressure of the electron gas as well as the field pressure parallel and perpendicular to the direction of magnetic field, are different. This creates anisotropy in the system. The total perpendicular pressure is greater than the total parallel pressure and hence such anisotropic nonrotating magnetized white dwarfs become deformed oblate spheroids. In order to calculate the masses and radii of these deformed axisymmetric white dwarfs we have considered a deformed Schwarzschild metric, known as $\gamma$-metric based on the deformation parameter $\gamma$ which is the ratio of the polar to equatorial radius. In order to incorporate $\gamma$ with the anisotropy of the EoS, we took $\gamma$ to be the ratio of the central parallel pressure to the central perpendicular pressure and keep it constant as done in \cite{Te19}. The magnitude of $\gamma$ is very small, so that we treat this deformation as a small perturbation to the spherically symmetric case. Moreover, we have taken a density-dependent magnetic field profile, so that the system is thermodynamically stable (both the parallel and perpendicular pressures are positive definite). We have solved the modified anisotropic stellar structure equations to obtain the masses and radii. First, we have computed the masses and equatorial radii with a fixed central electron density of $\sim 2.7 \times 10^{-6}$ $fm^{-3}$ and varying $B_{DC}$ values from 1 to 8 to show the effect of anisotropy due to magnetic field. Next we have computed the mass-central electron density, equatorial radius-central electron density and mass-equatorial radius relationships for $B_{DC}$ values from 1 to 8.
We found that for a fixed central electron number density, the mass decreases and the equatorial radius increases monotonically with increasing central magnetic field strength. This is because the total perpendicular pressure increases and the parallel pressure decreases resulting in more oblateness. As mentioned earlier, increasing magnetic field softens the EoS by decreasing $P_{avg}=(P_{\parallel T}+2P_{\perp T})/3$ which in turn decreases the mass. Due to the same reason, the maximum mass and its corresponding radius both shift at higher central electron densities as $B_{DC}$ increases. Both the values of the maximum mass and its corresponding radius decrease as $B_{DC}$ increases, implying greater compactification due to softening of the EoS. These results are very different from those of the isotropic case \cite{Mu17}, also we found that the maximum mass occurs for lower $B_{DC}$ at lower central electron number density and for higher $B_{DC}$, it occurs at higher central electron number density. \\
In this study, we have considered the anisotropy or deformation parameter $\gamma$ to be the ratio of the total central parallel pressure to the total central perpendicular pressure and hence is kept constant throughout the star. In reality, the anisotropy in the EoS is not constant but varies from the centre to the surface, due to the variation of the density and the magnetic field strength. We plan to incorporate the variable nature of the anisotropy parameter into the stellar structure equations in future. Nevertheless, our study shows that even for slight constant anisotropy inside the star, the gravitational mass can go to very high values, much larger than the Chandrasekhar limit. On the other hand, very high magnetic fields induce large anisotropy which leads to EoS softening and may lead to gravitational collapse. It would be interesting to apply this formalism to magnetars and magnetized quark stars, and also magnetized white dwarfs at finite temperature which we plan to work out in future.

\section{Acknowledgments}
The authors would like to acknowledge National Supercomputing Mission (NSM) for providing computing resources of Param Porul at NIT Trichy, which is implemented by C-DAC and supported by the Ministry of Electronics and Information Technology (MeitY) and Department of Science and Technology (DST), Government of India. The authors would also like to acknowledge Mr. Arunkarthiheyan Thiyagarajan for his valuable comments and suggestions.



\bibliographystyle{plainnat}
\bibliography{bibfile}
\end{document}